\documentclass[12pt,letterpaper,oneside,onecolumn]{article}
\usepackage[]{fontenc}
\usepackage[dvips]{epsfig}

\makeatletter
\def\LyX{L\kern-.1667em\lower.25em\hbox{Y}\kern-.125emX\spacefactor1000}%
\newcommand{\lyxtitle}[1] {\thispagestyle{empty}
\global\@topnum\z@
\section*{\LARGE \centering \sffamily \bfseries \protect#1 }
}

\makeatother

\begin{document}

\title{A Lax Representation for the Born-Infeld Equation}

\author{J. C. Brunelli\protect \thanks{e-mail address:brunelli@fsc.ufsc.br\protect }\\
Universidade Federal de Santa Catarina \\
Departamento de F\protect \'{\i}sica -- CFM\\
Campus Universit\protect \'{a}rio -- Trindade\\
C.P. 476, CEP 88040-900\\
Florian\protect \'{o}polis, SC -- BRAZIL\\
\\
and\\
\\
Ashok Das\\
Department of Physics and Astronomy\\
University of Rochester\\
Rochester, NY 14627 -- USA}

\date{~}

\maketitle

\begin{abstract}

We study the Born-Infeld equation from a Lagrangian point of view
emphasizing the duality symmetry present in such systems. We obtain
the Hamiltonian formulation directly from the Lagrangian. We also
show that this system admits a dispersionless nonstandard Lax representation
which directly leads to all the conserved charges (including the ones
not previously obtained). We also present the generating function
for these charges and point out various other properties of this system.

\end{abstract}

\pagebreak

Born and Infeld [1] in 1934 introduced a nonlinear generalization
of\break Maxwell's electrodynamics. Although their theory is not supported
by experiments, it can overcome the self-energy problem for the classical
electron. More recently, however, generalizations of Born-Infeld theory (BI)
have appeared in the context of effective actions following from the
string theory [2] as well as in the worldsheet actions for membranes [3].
In this letter, therefore, we want to study systematically this system
in \( 1+1 \) dimensions and try to connect various properties that are already
known about it. First, we study this system from a Lagrangian
point of view bringing out its duality invariance. This electric-magnetic duality (\( Z_{2} \) duality) plays a fundamental
role in the study of strings [4]. We, then, obtain the Hamiltonian formulation
of the system from the Lagrangian formulation making contact with
the earlier known results [5] for the BI equation both in
the physical as well as the null coordinates. We construct a dispersionless
nonstandard Lax representation for these equations which directly
gives the conserved charges of the system. This, of course, reproduces
the known conserved charges, but also gives a second
set of charges not known earlier for the BI system. We also present the generating function
for these charges from which many interesting properties
can be derived.

The simplest BI wave equation in \( 1+1 \)dimensions, that preserves Lorentz invariance and is nonlinear, is given in terms of a single scalar
field \( \phi  \) as

\begin{equation}
\label{A1}
(1+\phi _{x}^{2}\, )\phi _{tt}-2\phi _{x}\, \phi _{t}\, \phi _{xt}-(1-\phi _{t}^{2}\, )\phi _{xx}=0
\end{equation}
where the subscripts represent derivatives with respect to the appropriate
variables. This equation is integrable [6] and has a multi-Hamiltonain structure [5]. It is easy to check that the dynamical equation (\ref{A1}) can be
obtained from the Lagrangian density

\begin{equation}
\mathcal{L}=\sqrt{1-\phi _{t}^{2}+\phi _{x}^{2}}
\end{equation}
which leads to a Nambu-Goto type of action so relevant for a string
description. However, it is also easy to check that the dynamical
equation (\ref{A1}) can alternately be derived from the Lagrangian density
(Our metric is diagonal with the elements \( \eta ^{\mu \nu }=(+,-) \).)

\begin{equation}
\label{A2}
\mathcal{L}=J^{\mu }\, \partial _{\mu }\phi +\sqrt{1+J^{\mu }J_{\mu }}
\end{equation}
The Euler-Lagrange equations following from this Lagrangian are given
by

\[
\frac{J^{\mu }}{\sqrt{1+J^{\lambda }J_{\lambda }}}=-\partial ^{\mu }\phi \]
which we can equivalently write as
\begin{equation}
\label{A3}
J^{\mu }=-\frac{\partial ^{\mu }\phi }{\sqrt{1-\partial ^{\mu }\phi \, \partial _{\mu }\phi }}
\end{equation}
and

\begin{equation}
\label{A4}
\partial _{\mu }\, J^{\mu }=-\partial _{\mu }\, \left( \frac{\partial ^{\mu }\phi }{\sqrt{1-\partial ^{\mu }\phi \, \partial _{\mu }\phi }}\right) =0
\end{equation}
We note that the first equation (\ref{A3}) defines \( J^{\mu } \) in terms of the dynamical
field variables whereas the second (\ref{A4}) defines the dynamical equation
(\ref{A1}) as a current conservation.

Given the current \( J^{\mu } \), in \( 1+1 \) dimensions, we can define its dual as

\begin{equation}
\label{A5}
\tilde J ^{\mu }=-\epsilon ^{\mu \nu }J_{\nu }
\end{equation}
\\
where the Levi-Civita tensor is defined to be

\begin{equation}
\label{A6}
\epsilon ^{\mu \nu }=\left( \begin{array}{rc}
0 & 1\\
-1 & 0
\end{array}
\right) 
\end{equation}

Using eqs. (\ref{A3}) and (\ref{A5}), we can  also write

\begin{equation}
\label{A7}
\epsilon ^{\mu \nu }\partial _{\nu }\phi =\frac{\tilde J ^{\mu }}{\sqrt{1-\tilde J ^{\lambda }\tilde J _{\lambda }}}
\end{equation}
and it follows from this that

\begin{equation}
\label{A8}
\partial _{\mu }\, \left( \frac{\tilde J ^{\mu }}{\sqrt{1-\tilde J ^{\lambda }\tilde J_{\lambda }}}\right) =0
\end{equation}
\\
Comparing eqs. (\ref{A4}) and (\ref{A8}), we note that they go into each other under

\begin{equation}
\label{A9}
\partial ^{\mu }\phi \, \, \leftrightarrow \, \, \tilde J ^{\mu }
\end{equation}
\\
Namely, the dynamical equation in one variable goes into the Bianchi
identity in the dual variable. This is the duality invariance of the
dynamical equations and not of the Lagrangian which is the analogue
of the electric-magnetic duality in string theories.

The Lagrangian in eq. (\ref{A2}) is a first order Lagrangian and, consequently,
one can go to the Hamiltonian formulation in a straight forward manner
in various ways. We will use here the Faddeev-Jackiw formalism [7] for
simplicity. Writing out the Lagrangian density explicitly, we have

\begin{equation}
\label{A10}
\mathcal{L}=J_{t}\, \phi _{t}-J_{x}\, \phi _{x}+\sqrt{1+J_{t}^{2}-J_{x}^{2}}=J_{t}\, \phi _{t}-\Phi 
\end{equation}
Here, \( J_{t} \) and \( J_{x} \) represent the two components of the current vector
(and not the derivatives with respect to the variables) and it is
clear from the form of the Lagrangian (\ref{A10}) that of the three variables
(\( \phi  \), \( J_{t} \) and \( J_{x} \)), \( J_{t} \) is the momentum 
conjugate to \( \phi  \) and that \( J_{x} \) is a constrained
variable which can be eliminated through the equation

\begin{equation}
\label{A11}
\frac{\partial \Phi }{\partial J_{x}}=0
\end{equation}
\\
With a little bit of algebra, it can be seen to give

\begin{equation}
\label{A12}
J_{x}=-\sqrt{\frac{1+J_{t}^{2}}{1+\phi _{x}^{2}}}\phi _{x}
\end{equation}
\\
Substituting this back into the Lagrangian density in eq. (\ref{A10}), we
obtain the Lagrangian density in terms of the true dynamical variables
to be

\begin{eqnarray}
\mathcal{L} & = & J_{t}\, \phi _{t}+\sqrt{(1+\phi _{x}^{2})\, (1+J_{t}^{2})}=J_{t}\, \partial _{x}^{-1}(\phi _{x})_{t}+\sqrt{(1+\phi _{x}^{2})\, (1+J_{t}^{2})}\nonumber \\
\label{A13}  & = & J_{t}\, \partial _{x}^{-1}(\phi _{x})_{t}-H(\phi _{x},J_{t})
\end{eqnarray}
\\
If we treat \( \phi _{x} \) and \( J_{t} \) to be the dynamical variables, then, we recognize
that the Hamiltonian in (\ref{A13}) is the same as the one obtained in [5].
Furthermore, the Hamiltonian structure, for the system, can be read
out from the Lagrangian density in (\ref{A13}) to be

\begin{equation}
\label{A14}
\mathcal{D}=\left( \begin{array}{cc}
0 & \partial \\
\partial  & 0
\end{array}
\right) 
\end{equation}
so that the dynamical equations following from the Lagrangian can
now be written as

\begin{equation}
\label{A15}
\partial _{t}\, \left( \begin{array}{c}
\phi _{x}\\
J_{t}
\end{array}
\right) =\mathcal{D}\, \left( \begin{array}{c}
\frac{\delta H}{\delta \phi _{x}}\\
\\
\frac{\delta H}{\delta J_{t}}
\end{array}
\right) 
\end{equation}
which can be seen to be the same as the ones obtained in [5] with
the identifications \( \phi _{x}=r \) and \( J_{t}=-s \). It is worth noting here that although
the initial Lagrangian density in (\ref{A2}) was not invariant under the
duality transformations of eq. (\ref{A9}), the Lagrangian density of 
(\ref{A13}) as
well as the equations in (\ref{A15}) are manifestly duality invariant. (Remember
that \( J_{t}=-\tilde J _{x} \).) This is surprising since we noted earlier that the duality
is an on-shell symmetry. However, we can understand this by noting
that in obtaining the final form of the Lagrangian density, we have
indeed used the constraint equations without which duality would not
be manifest.

It was already pointed out in [5] that the BI system can be mapped
to the Chaplygin gas system through a redefinition of the dynamical
variables. In an earlier paper, we had shown how the polytropic gas
systems can be given a Lax representation [8]. This, then, allows us to
give a Lax representation for the BI system through a dispersionless
nonstandard Lax equation. Let us consider the Lax function (in the
phase space \( x \), \( p \))

\begin{equation}
\label{A16}
L=p^{-2}+\frac{rs}{\sqrt{(1+r^{2})(1+s^{2})}}+\frac{p^{2}}{4(1+r^{2})(1+s^{2})}
\end{equation}
where we have used the variables of [5] for ease of comparison. Expanding
the square root of this function around \( p=0 \) (See ref. [8] for details.), we obtain

\begin{equation}
\label{A17}
\left( L^{{1}/{2}}\right) _{\leq 1}=\frac{rs}{2\sqrt{(1+r^{2})(1+s^{2})}}p+p^{-1}
\end{equation}
It can, then be easily checked that the dynamical equations (\ref{A15}) can
be written in the Lax form

\begin{equation}
\label{A18}
\frac{\partial L}{\partial t}=2\left\{\left( L^{{1}/{2}}\right)_{\leq 1},L\right\}
\end{equation}
\\

The Lax description is quite useful in obtaining various properties
about the system. For example, we immediately recognize that the conserved
quantities associated with the system can be obtained from the residues
of  various powers of the Lax function. We note that if we expand
the square root around \( p=0 \), as we have done earlier, we can obtain a
set of conserved densities associated with the system to be

\begin{equation}
\label{A19}
H_{n-1}=\frac{2^{n}}{(2n+1)!!}\, \hbox{Res}L^{\frac{2n+1}{2}}\,,\qquad n=1,2,3,\dots 
\end{equation}
Explicitly, this gives the first few Hamiltonian densities to be

\begin{eqnarray}
H_{0} & = & \frac{rs}{\sqrt{(1+r^{2})(1+s^{2})}}\nonumber \\
H_{1} & = & \frac{1+3r^{2}s^{2}}{6(1+r^{2})(1+s^{2})}\nonumber \\
\label{A20} H_{2} & = & \frac{rs(1+r^{2}s^{2})}{6(1+r^{2})^{3/2}(1+s^{2})^{3/2}}
\end{eqnarray}

Alternately, we could have expanded the square root of the Lax function
around \( p=\infty  \). This would lead to a second set of charges that are conserved.
Defining, the Hamiltonian densities

\begin{equation}
\label{A21}
\tilde H _{n}=\frac{2^{n}}{(2n+1)!!}\,\hbox{Res} L^{\frac{2n+1}{2}}\,,\qquad n=-1,0,1,\dots 
\end{equation}
We can now obtain the first few conserved densities from this to
be

\begin{eqnarray}
\tilde H _{-1} & = & \sqrt{(1+r^{2})(1+s^{2})}\nonumber \\
\tilde H _{0} & = & rs\nonumber \\
\tilde H _{1} & = & \frac{1+r^{2}s^{2}}{2\sqrt{(1+r^{2})(1+s^{2})}}\nonumber \\
\label{A22} \tilde H _{2} & = & \frac{rs(3+r^{2}s^{2})}{6(1+r^{2})(1+s^{2})}
\end{eqnarray}
We recognize the first three as the conserved densities obtained
in [5]. However, the set of densities in eq. (\ref{A20}) are new. In
fact, since we have the Lax representation for the equation, we can
also construct the generating function for the conserved densities,
following the method described in [8]. Without going into details,
let us simply note here that the generating function for the two sets
of Hamiltonian densities are given by

\begin{equation}
\label{A23}
\chi =\left[ -\frac{rs+\lambda \sqrt{(1+r^{2})(1+s^{2})}}{2\sqrt{(1+r^{2})(1+s^{2})}}+\left\{ \frac{\left( rs+\lambda \sqrt{(1+r^{2})(1+s^{2})}\right) ^{2}-1}{4(1+r^{2})(1+s^{2})}\right\} ^{\frac{1}{2}}\right] ^{-\frac{1}{2}}
\end{equation}

\begin{eqnarray}
\tilde\chi  & = & \left[ \frac{rs+\lambda \sqrt{(1+r^{2})(1+s^{2})}}{2\sqrt{(1+r^{2})(1+s^{2})}}+\left\{ \frac{\left( rs+\lambda \sqrt{(1+r^{2})(1+s^{2})}\right) ^{2}-1}{4(1+r^{2})(1+s^{2})}\right\} ^{\frac{1}{2}}\right] ^{-\frac{1}{2}}\nonumber 
\end{eqnarray}
Following [8], it can be shown that both \( \chi  \) and \( \tilde\chi  \) are conserved for
any value of \( \lambda  \) and consequently, expanding these around \( \lambda =\infty  \), one recovers
the two sets of conserved quantities. We can also show from the generating
functions (following [8]) that the conserved charges are in involution. Furthermore, we can also read
out from the forms of the generating functions that the Riemann invariants
associated with the BI equation are given by

\begin{equation}
\label{A23a}
-\lambda _{\pm }=\frac{rs\pm 1}{\sqrt{(1+r^{2})(1+s^{2})}}
\end{equation}

It is worth noting here that if we expand the square root of \( L \) around
\( p=\infty  \), we obtain

\[
\left(L^{{3}/{2}}\right)_{\geq 2}=\frac{p^{3}}{8(1+r^{2})^{3/2}(1+s^{2})^{3/2}}\]
and this leads to the consistent Lax equation

\[
\frac{\partial L}{\partial t}=\left\{ \left( L^{{3}/{2}}\right)_{\geq 2},L\right\} \]
which can be seen to be equivalent to the dynamical equations

\[
r_{t}=\frac{3s_{x}}{4(1+s^{2})^{2}}\]

\begin{equation}
\label{A24}
s_{t}=\frac{3r_{x}}{4(1+r^{2})^{2}}
\end{equation}
We do not know whether this equation has been studied in the past,
but it is clear from the Lax representation that this equation is
also integrable and shares the same set of conserved charges with
the BI system. We note here that with \( H=\frac{3}{4}\sqrt{(1+r^{2})(1+s^{2})} \), the set of equations (\ref{A24})
can be written in the Hamiltonian form

\[
\partial _{t}\, \left( \begin{array}{c}
r\\
s
\end{array}
\right) =\left( \begin{array}{cc}
0 & \frac{1}{\sqrt{(1+s^{2})}}\partial \frac{1}{\sqrt{(1+r^{2})}}\\
\frac{1}{\sqrt{(1+r^{2})}}\partial \frac{1}{\sqrt{(1+s^{2})}} & 0
\end{array}
\right) \left( \begin{array}{c}
\frac{\delta H}{\delta r}\\
\\
\frac{\delta H}{\delta s}
\end{array}
\right) \]
We have not, however, checked whether this is a genuine Hamiltonian
structure in the sense whether it satisfies the Jacobi identity.

It was pointed out in ref. [5] that the BI equation, under a redefinition
of variables goes into the Chaplygin gas equation. It is worth analyzing
this question in some detail since it appears to be a general property
of such systems. We note that, for such first order systems of Hydrodynamic
type, the dynamical equations have the form of a conservation law.
The conserved densities of such systems also satisfy a first order
equation. Consequently, in such systems, it is possible to define
new variables which correspond to some of the conserved densities
of the system and thereby go to other first order systems of Hydrodynamic
type. Thus, for example, if we were to choose

\begin{equation}
\label{A25}
u=H_{0}=\frac{rs}{\sqrt{(1+r^{2})(1+s^{2})}}\, \, \, \, \, \, \, \, \, \, \, \, \, \, \, \, \, \, \, \, \, v=\tilde H _{-1}=\sqrt{(1+r^{2})(1+s^{2})}
\end{equation}
then, it is easy to check that they satisfy

\begin{eqnarray}
u_{t} & = & -uu_{x}-v^{-3}v_{x}\nonumber \\
\label{A26} v_{t} & = & -(uv)_{x}
\end{eqnarray}
which are the Chaplygin gas equations. We emphasize that such a field
redefinition is not a point transformation in the sense that it does
not leave the Hamiltonian structure invariant even though both the
BI equation as well as the Chaplygin gas equation do share the same
Hamiltonian structure of eq. (\ref{A14}). Furthermore, this transformation
neither leaves the Hamiltonian of the BI equation invariant nor does
it take the BI Hamiltonian to that of the Chaplygin gas Hamiltonian.
Rather, the BI Hamiltonian, under the field redefinition, goes into
one of the infinite conserved quantities of the system. This is
 a general property of first order systems of Hydrodynamic type.
In fact, under the field redefinitions only the Riemann invariants
go into each other as can be checked from the fact that the Riemann
invariants for the Chaplygin gas equations are given by

\begin{equation}
\label{A26a}
-\lambda _{\pm }=u\pm \frac{1}{v}
\end{equation}
which go under the field redefinitions of eq. (\ref{A25}) into the Riemann
invariants of the BI equation, namely, (\ref{A23a}). As a second example, we
note that the equations (\ref{A24}) under the same field redefinitions (\ref{A25})
can be easily checked to go into the elastic medium equations

\begin{eqnarray}
u_{t} & = & -\frac{1}{4}\left( \frac{1}{v^{3}}\right) _{x}\nonumber \\
\label{A27} v_{t} & = & \frac{3}{4}u_{x}
\end{eqnarray}
which are known to have the same Riemann invariants as the Chaplygin
gas equation.

We can also consider the BI equation in the null coordinates in the
same spirit. Without going into too much detail, let us note some
of the features of this system here. Defining the light cone coordinates

\begin{eqnarray}
t'&=& \frac{1}{2}(t+x)\nonumber \\
\label{A28} x'&=& \frac{1}{2}(t-x)
\end{eqnarray}
under which the metric and the Levi-Civita tensor transform as

\begin{eqnarray}
\eta ^{\mu \nu } & \rightarrow &\eta '^{\, \mu \nu }=\frac{1}{2}\left( \begin{array}{cc}
0 & 1\\
1 & 0
\end{array}
\right) \nonumber \\
\label{A29} \epsilon ^{\mu \nu } & \rightarrow & \epsilon '^{\, \mu \nu }=\frac{1}{2}\left( \begin{array}{cr}
0 & -1\\
1 & 0
\end{array}
\right) 
\end{eqnarray}
It is easy to see that under such a transformation, the BI equation
of (\ref{A1}) changes to

\begin{equation}
\label{A30}
\phi _{x'}^{2}\, \phi _{t't'}+2(2+\phi _{t'}\, \phi _{x'}\, )\phi _{t'x'}+\phi _{t'}^{2}\, \phi _{x'x'}=0
\end{equation}
This is the BI equation in the null coordinates. Once again this
can be derived from the Lagrangian density of (\ref{A2}), which can be written
in the new coordinates as

\begin{eqnarray}
\mathcal{L} & =&J^{\mu }\partial _{\mu }\phi +\sqrt{1+J^{\mu }J_{\mu }}\nonumber \\
\label{A31}  &  =&\frac{1}{2}(J_{x'}\, \phi _{t'}+J_{t'}\, \phi _{x'})+\sqrt{(1+J_{t'}\, J_{x'})}=\frac{1}{2}J_{x'}\, \phi _{t'}-\Phi 
\end{eqnarray}
It is clear, as before, that of the three dynamical variables, \( J_{x'} \)
is the momentum conjugate to \( \phi  \) and that \( J_{t'} \) is a constrained variable
which can be eliminated through the equation

\[
\frac{\partial \Phi }{\partial J_{t'}}=0\]
This, with some algebra, can be seen to lead to

\[
J_{t'}=\frac{J_{x'}}{\phi _{x'}^{2}}-\frac{1}{J_{x'}}\]
Substituting this back into the Lagrangian density (\ref{A31}), gives

\begin{equation}
\label{A32}
\mathcal{L}=\frac{1}{2}\left[ J_{x'}\, \partial _{x'}^{-1}(\phi _{x'})_{t'}-\left( \frac{\phi _{x'}}{J_{x'}}+\frac{J_{x'}}{\phi _{x'}}\right) \right] =\frac{1}{2}\left( J_{x'}\, \partial _{x'}^{-1}(\phi _{x'})_{t'}-H\right) 
\end{equation}
Once again, the dynamical equations can be obtained from this Lagrangian
density which coincide with the null equations of ref. [5] with the
identifications \( \phi _{x'}=w \) and \( J_{x'}=z \). The Hamiltonian structure can also be read
out from eq. (\ref{A32}) and coincides with (\ref{A14}). Similarly, the Hamiltonian
also coincides with the one obtained in [5]. The duality invariance
(\( \phi _{x'}\, \leftrightarrow \, J_{x'} \) or \( w\leftrightarrow z \)) is again manifest here both in the Lagrangian density as well
as the equations of motion.

The BI equation in the null coordinates can again be given a Lax representation
and we simply note here that the Lax function (once again we use the
variables of [5] for ease of comparison)

\begin{equation}
\label{A33}
L=p^{-2}+\left( \frac{1}{w^{2}}+\frac{1}{z^{2}}\right) +\frac{1}{w^{2}z^{2}}p^{2}
\end{equation}
gives the null coordinate equations through the nonstandard Lax equation

\[
\frac{\partial L}{\partial t'}=2\left\{(L^{{1}/{2}})_{\leq 1}, L\right\}\]
when the square root is expanded around \( p=0 \). As before, we can construct
the conserved quantities from this Lax function and there are two
sets of conserved densities from \( \hbox{Res}\,L^{(2n+1)/2} \). Expanding the square root around
\( p=0 \), leads to the densities of the form (with appropriate normalizations)

\begin{eqnarray}
H_{0} &   =&\frac{1}{w^{2}}+\frac{1}{z^{2}}\nonumber \\
H_{1} &  =&\frac{1}{w^{4}}+\frac{10}{3w^{2}z^{2}}+\frac{1}{z^{4}}\nonumber \\
H_{2} &  =&\frac{1}{w^{6}}+\frac{7}{w^{4}z^{2}}+\frac{7}{w^{2}z^{4}}+\frac{1}{z^{6}}\nonumber \\
\label{A34} H_{3} &   =&\frac{1}{w^{8}}+\frac{12}{w^{6}z^{2}}+\frac{126}{5w^{4}z^{4}}+\frac{12}{w^{2}z^{6}}+\frac{1}{z^{8}}
\end{eqnarray}
while expansion of the square root around \( p=\infty  \), leads to the conserved
densities of the form

\begin{eqnarray}
\tilde H _{-1} &=& wz\nonumber \\
\tilde H _{0} &=& \frac{w}{z}+\frac{z}{w}\nonumber \\
\tilde H _{1} &=& \frac{w}{z^{3}}+\frac{6}{wz}+\frac{z}{w^{3}}\nonumber \\
\label{A35} \tilde H _{2} & =& \frac{w}{z^{5}}+\frac{15}{w^{3}z}+\frac{15}{wz^{3}}+\frac{z}{w^{5}}
\end{eqnarray}
Once again, we note here that the second set of densities in (\ref{A35})
are already obtained in [5]. However, the set of densities in (\ref{A34}) are
new. Without going into details, let us note here that the generating
function for these charges can be obtained to be

\begin{eqnarray*}
\chi  &=&\left[ \frac{1}{2}\left\{ \left( \frac{1}{w^{2}}+\frac{1}{z^{2}}+\lambda \right) ^{2}-\frac{4}{w^{2}z^{2}}\right\} ^{\frac{1}{2}}-\frac{1}{2}\left( \frac{1}{w^{2}}+\frac{1}{z^{2}}+\lambda \right) \right] ^{-\frac{1}{2}}\\
\tilde \chi  & =&\left[ \frac{1}{2}\left\{ \left( \frac{1}{w^{2}}+\frac{1}{z^{2}}+\lambda \right) ^{2}-\frac{4}{w^{2}z^{2}}\right\} ^{\frac{1}{2}}+\frac{1}{2}\left( \frac{1}{w^{2}}+\frac{1}{z^{2}}+\lambda \right) \right] ^{-\frac{1}{2}}
\end{eqnarray*}
>From this, we can also see now that the Riemann invariants of the
BI equations in the null coordinates are given by

\begin{equation}
\label{A36}
-\lambda _{\pm }=\left( \frac{w\mp z}{wz}\right) ^{2}
\end{equation}
from which we can obtain

\begin{eqnarray}
w &=&\frac{2}{\sqrt{-\lambda _{-}}+\sqrt{-\lambda _{+}}}\nonumber \\
\label{A37} z &  =&\frac{2}{\sqrt{-\lambda _{-}}-\sqrt{-\lambda _{+}}}
\end{eqnarray}
It now follows from eqs. (\ref{A23a}), (\ref{A36}) and (\ref{A37}) that the field redefinitions
which will take the general BI equation to the one in the null coordinates
have to be

\begin{eqnarray}
w &  =&(1+r^{2})^{1/4}(1+s^{2})^{1/4}[(rs+1)^{1/2}-(rs-1)^{1/2}]\nonumber \\
\label{A38} z & =&(1+r^{2})^{1/4}(1+s^{2})^{1/4}[(rs+1)^{1/2}+(rs-1)^{1/2}]
\end{eqnarray}
These are indeed the field redefinitions obtained in [5]. Our analysis
clarifies how they arise. Also, the analogue of (27) is
\begin{equation}
\label{bru1}
u=H_{0}=\frac{1}{w^2}+\frac{1}{z^2}\, \, \, \, \, \, \, \, \, \, \, \, \, \, \, \, \, \, \, \, \, v=\tilde H _{-1}=wz
\end{equation}
This, with appropriate redefinitions, is the Verosky transformation used in ref. [5] and it can be shown to satisfy the Chaplygin gas equation (28).

Finally, we would like to point out that the Chaplygin gas like equations
\begin{eqnarray}
u_t+uu_x+{v_x\over v^{\alpha+2}}&=&0\,,\quad \alpha\ge1\nonumber \\
\label{bru2}v_t+(uv)_x&=&0
\end{eqnarray}
have a Lax representation of the form
\begin{equation}
\label{bru3}
{\partial L\over\partial t}={(\alpha+1)\over\alpha}
\left\{\left(L^{\alpha\over\alpha+1}\right)_{\le1},L\right\}
\end{equation}
with the Lax function
\begin{equation}
\label{bru4}
L=p^{-(\alpha+1)}+u+{v^{-(\alpha+1)}\over(\alpha+1)^2}\,p^{\alpha+1}\,,\quad\quad \alpha\ge1
\end{equation}
Here,  $L^{1\over\alpha+1}$ is expanded around $p=0$. Of course, (28) is the special case $\alpha=1$. Conserved densities associated with (43) can be obtained from
\begin{equation}
\label{bru5}
H_n=\hbox{Res}\,L^{n+{\alpha+2\over\alpha+1}},\qquad n=0,1,2,3,\dots
\end{equation}
expanding $L^{1\over\alpha+1}$ around
$p=0$, and from
\begin{equation}
\widetilde{H}_n=\hbox{Res}\,L^{n-{1\over\alpha+1}},\quad n=0,1,2,3,\dots
\end{equation}
expanding $L^{1\over\alpha+1}$ around $p=\infty$.
This complements our results in [8] where a Lax description for the polytropic gas equations (eq. (43) with $\alpha\le2$) was obtained.

\vspace{.5cm}
\noindent{\bf Note Added:} Since the paper was submitted, we have become aware of the interesting family of Monge-Amp\`ere equations [9-11] some of which may be related to the Born-Infeld equations. (We would like to thank the referee for bringing this to our attention.) For example, the hyperbolic Monge-Amp\`ere equation
\begin{equation}
U_{tt}U_{xx}-(U_{tx})^2=-1
\end{equation}
expressed in terms of the first order variables [9]
\begin{equation}
A=U_x\qquad B=U_t
\end{equation}
can be related to the Chaplygin gas equation (see eqs. (28),(43)) through the transformation [10]
\begin{equation}
u=-{B_x\over A_x}\qquad v=A_x
\end{equation}
Thus, we can even give a Lax description for the hyperbolic Monge-Amp\`ere equation through (see eqs. (44)-(45))

\begin{eqnarray}
{\partial L\over\partial t}& = &2
\left\{\left(L^{1/2}\right)_{\le1},L\right\}\nonumber \\
L&=&p^{-2}-{B_x\over A_x}+{p^2\over 4A_x^2}\nonumber 
\end{eqnarray}
Of course, the Chaplygin gas equations are related to the Born-Infeld equations. It follows now that all our discussions can be carried over to this class of systems as well.

\vspace{.5cm}
J.C.B. was supported by CNPq, Brazil. A.D was supported by the US
Department of Energy Grant No. DE-FG-02-91ER40685 and by NSF-INT-9602559.

\vspace{.5cm}
\vfill\eject
\begin{large}
\noindent{\bf References:}
\end{large}

\vspace{.5cm}
\begin{enumerate}
\item{ }M. Born and L. Infeld, Proc. R. Soc. London {\bf A144}, 425 (1934).
\item{ }E. S. Fradkin and A. A. Tseytlin, Phys. Lett. {\bf B163}, 123 (1985).
\item{ }J. Polchinski, ``Tasi Lectures on D-branes'', hep-th/9611050; M. \break Aganagic, J. Park, C. Popescu and J. Schwartz, Nucl. Phys. {\bf B496}, 191 (1997). 
\item{ }G. W. Gibbons and D. A. Rasheed, Nucl. Phys. {\bf B454}, 185 (1995); Phys. Lett. {\bf B365}, 46 (1996).
\item{ } M. Arik, F. Neyzi, Y. Nutku, P. J. Olver and J. Verosky, J. Math. Phys. {\bf 30}, 1338 (1988).
\item{ }B. M. Barbishov and N. A. Chernikov, Sov. Phys. JETP {\bf 24}, 93 (1966).
\item{ }L. Faddeev and R. Jackiw, Phys. Rev. Lett. {\bf 60}, 1692 (1988); J. Barcelos Neto and C. Wotzasek, Int. J. Mod. Phys. {\bf A7}, 4981 (1992).
\item{ }J. C. Brunelli and A. Das, Phys. Lett. {\bf A235}, 597 (1997). 

\item{ } Y. Nutku and \"O. Sario\v glu, Phys. Lett. {\bf A173}, 270 (1993); Y. Nutku, J. Phys. {\bf 29A}, 3257 (1996) 

\item{ } O.I. Mokhov and Y. Nutku, Lett. Math. Phys. {\bf 32}, 121 (1994).

\item{ } K. J\"orgens, Math. Annal. {\bf 127}, 130 (1954); E. Heinz, Nach. Akad. Wissensch. in G\"ottingen Mathem.-Phys. Klasse {\bf IIa}, 51 (1952).
\end{enumerate}

\end{document}